\documentclass[aps,prl,reprint,nofootinbib,superscriptaddress]{revtex4-2}

\usepackage{graphicx}
\usepackage{amsmath}
\usepackage{amssymb}
\usepackage{hyperref}
\usepackage{color}

\DeclareMathOperator{\Imag}{Im}

\begin{document}

\title{Anyon braiding on a fractal lattice with a local Hamiltonian}

\author{Sourav Manna}
\affiliation{Max-Planck-Institut f\"ur Physik komplexer Systeme, D-01187 Dresden, Germany}
\affiliation{Department of Condensed Matter Physics, Weizmann Institute of Science, Rehovot 7610001, Israel}

\author{Callum W. Duncan}
\affiliation{Max-Planck-Institut f\"ur Physik komplexer Systeme, D-01187 Dresden, Germany}
\affiliation{Department of Physics, SUPA and University of Strathclyde, Glasgow G4 0NG, United Kingdom}

\author{Carrie A. Weidner}
\affiliation{Department of Physics and Astronomy, Aarhus University, DK-8000 Aarhus C, Denmark}

\author{Jacob F. Sherson}
\affiliation{Department of Physics and Astronomy, Aarhus University, DK-8000 Aarhus C, Denmark}

\author{Anne E. B. Nielsen}
\affiliation{Max-Planck-Institut f\"ur Physik komplexer Systeme, D-01187 Dresden, Germany}
\affiliation{Department of Physics and Astronomy, Aarhus University, DK-8000 Aarhus C, Denmark}

\begin{abstract}
There is a growing interest in searching for topology in fractal dimensions with the aim of finding different properties and advantages compared to the integer dimensional case. It has previously been shown that the Laughlin state can be adapted to fractal lattices. A key element in doing so is to replace the uniform background charge by a background charge that resides only on the lattice sites. This motivates the study of Hofstadter type models on fractal lattices, in which the magnetic field is present only at the lattice sites. Here, we study such models for hardcore bosons on finite lattices derived from the Sierpinski carpet and on square lattices with open boundary conditions. We find that the system sizes that we can investigate with exact diagonalization are generally too small to judge whether these local models are topological or not. Studying the particle densities on the lattices derived from the Sierpinski carpet, we find that the densities tend to accumulate in the regions that are locally similar to a square lattice. Such accumulation seems to be incompatible with the uniform densities in fractional quantum Hall systems, which might suggest that the models are not topological. Our computations provide guidance for future searches for topology in finite systems. We also propose a scheme to implement both fractal lattices and our proposed local Hamiltonian with ultracold atoms in optical lattices, which could allow for quantum simulators to go beyond the numerical results presented here.
\end{abstract}

\maketitle

\let\thefootnote\relax\footnotetext{The paper uses figures, video, table, and modified text from the papers [Anyon braiding on a fractal lattice with a local Hamiltonian, S. Manna, C. W. Duncan, C. A. Weidner, J. Sherson \& A. E. B. Nielsen, Phys.\ Rev.\ A \textbf{105}, L021302 (2022), DOI:10.1103/PhysRevA.105.L021302] and [Erratum: Anyon braiding on a fractal lattice with a local Hamiltonian [Phys. Rev. A 105, L021302 (2022)], S. Manna, C. W. Duncan, C. A. Weidner, J. Sherson \& A. E. B. Nielsen, Phys.\ Rev.\ A \textbf{107}, 069901 (2023), DOI:10.1103/PhysRevA.107.069901] under the \href{https://creativecommons.org/licenses/by/4.0/}{Creative Commons Attribution 4.0 International license}.}

Topologically ordered quantum systems harbor fractionalized excitations that are neither fermions nor bosons, but anyons \cite{Leinaas,wilczek}. Phases hosting anyons have been realized experimentally in solid state systems in strong magnetic fields displaying the fractional quantum Hall effect \cite{stormer,laughlin,arovas,bolotin}. Fractional quantum Hall phases also exist in systems defined on two-dimensional lattices, where the physical magnetic field is replaced by an artificial magnetic field, which can be much stronger \cite{Lukin2,Lukin1,kapit1,Bernevig1,Mudry1}. Due to their unique degree of tunability, realizing fractional quantum Hall physics with ultracold atoms in optical lattices would give unique possibilities for investigating the effect in great detail, and there are currently several efforts towards achieving this for systems with few atoms \cite{Dalibard1,adiabatic1,adiabatic2,Eckardt1,repellin}. The key components of artificial magnetic fields and topological band structures have already been prepared in several experiments \cite{reviewCooper}.

Topological phases are mainly studied in systems with spatial (and Hausdorff) dimension one, two, and three, but recently interest has grown in studying topological models on fractal lattices with non-integer Hausdorff dimension. The Hausdorff dimension is a generalisation of the dimension of a vector space and can provide a measure of how the details of a system change at different scales. While much of the knowledge generated in condensed matter physics relies on the presence of an underlying Bravais lattice, fractal lattices do not fit into this framework and can hence give rise to different physics. Most of the studies of topological quantum models on fractal lattices so far have considered non-interacting systems \cite{titus2,fritz1,pai2019topological,yuan1} and those have, indeed, revealed new and interesting properties, including modifications of the Hofstadter butterfly and the presence of inner edge states. Much less is currently known about how fractal lattices affect the properties of topologically ordered phases of interacting systems. Initial steps have been taken by constructing Laughlin and Moore-Read trial states on fractal lattices \cite{nielsen2,nielsen1}, but the derived parent Hamiltonians of these states are nonlocal and involve many different types of interactions making them difficult to realize. The study of models on fractal lattices is also motivated by experimental developments, such as the preparation of fractal models in molecules on surfaces \cite{shang2015assembling,kempkes2019design}. It is desirable to realize fractal models of matter with ultracold atoms due to their ability to reach the regime of strongly interacting quantum systems and achieve single-site resolution \cite{Greiner_QGM,Sherson_QGM}.

Fractional quantum Hall models on two-dimensional square lattices can be constructed by considering hardcore bosons hopping between nearest-neighbor sites in the presence of a uniform magnetic field \cite{Lukin2,Lukin1}. Here, we investigate whether nearest-neighbor hopping Hamiltonians for hardcore bosons can realize fractional quantum Hall models on small lattices obtained from fractals of finite generation. To adapt the Laughlin state to fractal lattices, the background charge is restricted to be on the lattice sites only, and since the background charge is proportional to the magnetic field, it is natural to consider models, in which the magnetic field goes through the lattice sites only. Such a magnetic field is implemented into the model through the Peierls substitution \cite{peierls1933theorie}.

Fractal lattices have open boundary conditions, which makes it harder to find suitable diagnostics for topology. One option is to demonstrate anyonic exchange statistics. We first consider the model on a square lattice. We add two pinning potentials and reduce the number of particles such that the total charge of the system is compatible with the presence of two quasiholes. In a topological system, one expects each potential to trap a charge of a particular value in a local region, if the potentials are sufficiently far apart to ensure that the regions do not overlap. While we do see accumulation of charge near the potentials, as long as the potentials are not too close to the edge, the systems we can investigate with exact diagonalization are not large enough to avoid overlap. When we adiabatically exchange the pinning potentials along a path that does not enclose an area, the phase acquired by the wavefunction varies by more than one half in units of $\pi$, depending on the precise choice of the potentials. Whether anyonic or trivial braiding statistics is obtained for systems that are large enough to avoid overlap is, however, unclear.

For the model on the Sierpinski carpet, we find that the particles in the ground state preferentially localize to domains that are locally similar to a two-dimensional square lattice. This seems incompatible with the uniform density of the bulk of fractional quantum Hall states and suggests that the local model we consider on the Sierpinski carpet may not be topological. Our computations provide guidance for future searches for topology in small lattices. This topic is of current relevance given the recent experimental progress in the area \cite{leonard2022realization}.

We also propose a scheme to implement the Hamiltonian experimentally with ultracold atoms in optical lattices. The proposed protocol involves single-site addressing \cite{Weitenberg_spin_addressing} and laser-assisted hopping tuned to achieve the desired phase factors \cite{Aidelsberger_Hofstadter}. We expect that the setup could also be used to generate integer quantum Hall phases on fractal lattices. Generating optical, fractal lattices as described below also opens the door for studying various phenomena of quantum systems on fractal lattices.

\begin{figure*}
\includegraphics[width=0.95\textwidth]{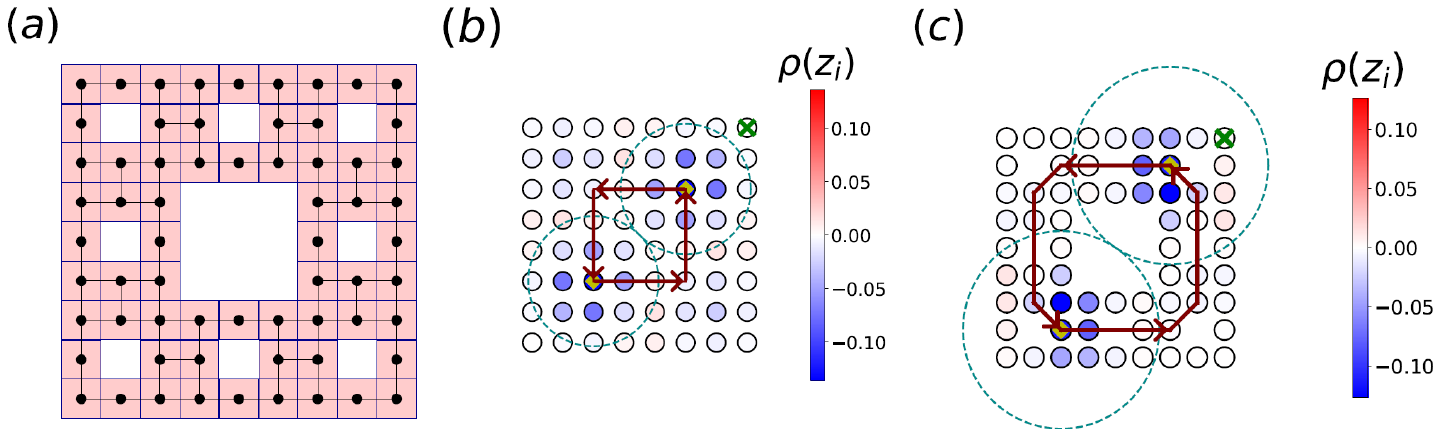}
\caption{(a) The model is defined on a second generation Sierpinski carpet (red squares). The lattice sites are marked by circles, and the bonds connecting the sites illustrate the hopping terms. A magnetic flux goes through each lattice site in the direction perpendicular to the plane. (b-c) The density profile $\rho(z_i)$ from \eqref{density} for the square and fractal lattices, when the trapping potentials are on the sites marked by diamonds. The dashed lines encircle the regions that we sum over when we compute the trapped charges. The arrows show a potential choice of exchange path. The trapped charges when one pinning potential is placed on the corner site (marked with the green cross) are computed in Fig.\ \ref{fig:density}.} \label{DensityChargeSqSC}
\end{figure*}

\textit{Model}---%
Typical ingredients required to obtain fractional quantum Hall physics include interactions and a magnetic field perpendicular to the plane. In lattice systems, the magnetic field is often translated into corresponding complex hopping terms through the Peierls substitution \cite{peierls1933theorie} as we shall also do below. We start from a lattice with $N$ sites embedded in two dimensions, such as the lattice in Fig.\ \ref{DensityChargeSqSC}(a), which is obtained from the second generation Sierpinski carpet. We denote the positions of the lattice sites in the complex plane by $z_j$ with $j\in\{1,\ldots,N\}$ and consider a fixed number $M$ of bosons on the lattice. The Hamiltonian
\begin{equation}\label{FCI}
H=-J\sum_{\langle jk \rangle} c_j^\dagger c_k e^{i \phi_{jk}}
+U\sum_l n_l(n_l-1), \quad U \gg J,
\end{equation}
consists of complex, nearest-neighbor hopping terms of strength $J$ and an on-site interaction term of strength $U$. The operator $c_k$ annihilates a boson on the $k$th lattice site, $n_k=c_k^{\dag} c_k$, and $\phi_{jk}$ is the phase the wavefunction acquires when a particle hops from $z_k$ to $z_j$. In the computations below, which are all done using exact diagonalization, we assume that $U/J$ is so large that one can neglect the possibility to have more than one boson on a site, i.e.\ we work with hardcore bosons.

The particular form of $\phi_{jk}$ is determined from the chosen magnetic field. In two-dimensional fractional quantum Hall models, the magnetic field is often either uniform or only penetrates the lattice sites. For a fractal lattice, it is similarly natural to let the magnetic field only penetrate the lattice sites, since then the pattern of magnetic flux also forms a fractal. We hence choose the magnetic field to be $\vec{B}(z)=\alpha\sum_l\delta(z-z_l)\hat{z}$, where $\alpha$ is the flux penetrating one lattice site measured in terms of the magnetic flux unit, $\delta$ is the Dirac delta function, and $\hat{z}$ is a unit vector perpendicular to the plane. This field configuration gives rise to the vector potential
\begin{equation}
\vec{A}(z)=\sum_l \frac{\alpha \hat{\theta}_l}{|z-z_l|},
\end{equation}
where $\hat{\theta}_l$ is a unit vector in the plane rotated by $\pi/2$ compared to $z-z_l$. From this we obtain
\begin{equation}
\phi_{jk}=\int_{z_k}^{z_j} \vec{A}(r) \cdot\vec{dl} = \alpha \sum_{l(\neq j\neq k)} \Imag\left[\ln\left(\frac{z_j-z_l}{z_k-z_l}\right)\right],
\end{equation}
where $\vec{dl}$ is an infinitesimal vector along the hopping direction. Below, we take $M/(\alpha N) = 1/2$, where $M$ is the number of particles. If the system is topological, we hence expect it to be in a bosonic Laughlin phase with quasiholes of charge $1/2$.

The model described above can also be defined on a square lattice, which we will do for comparison. We will consider open boundary conditions for both the square and fractal lattice for appropriate comparison. Note, periodic boundary conditions are not consistent with the fractal retaining its scaling nature.

\textit{Energy gap}---%
The energy gap, $\delta E$, between the ground state and the first excited state is an important property of the model. This is due to the gap's relation to the state's stability, both in terms of robustness to disorder and feasibility of experimental implementations. We find that the size of the energy gap varies substantially with the number of particles and is particularly large for the fractal lattice with $4$ particles. For $4$ particles, we find that the energy gap is about three times larger for the fractal lattice than for the square lattice. Specifically, the gap is $\delta E=0.313J$ for the fractal lattice and $\delta E=0.105J$ for the square lattice.

\begin{figure*}
{\includegraphics[width=0.32\linewidth]{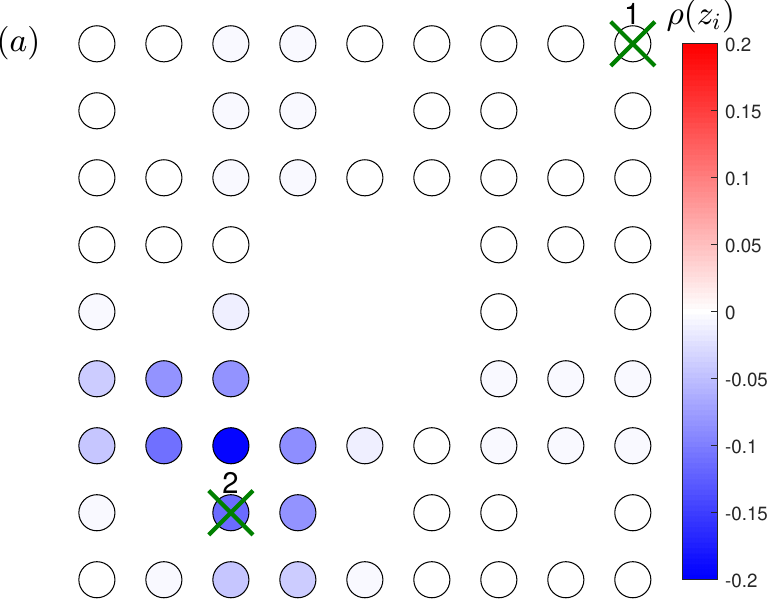}\hfill
\includegraphics[width=0.32\linewidth]{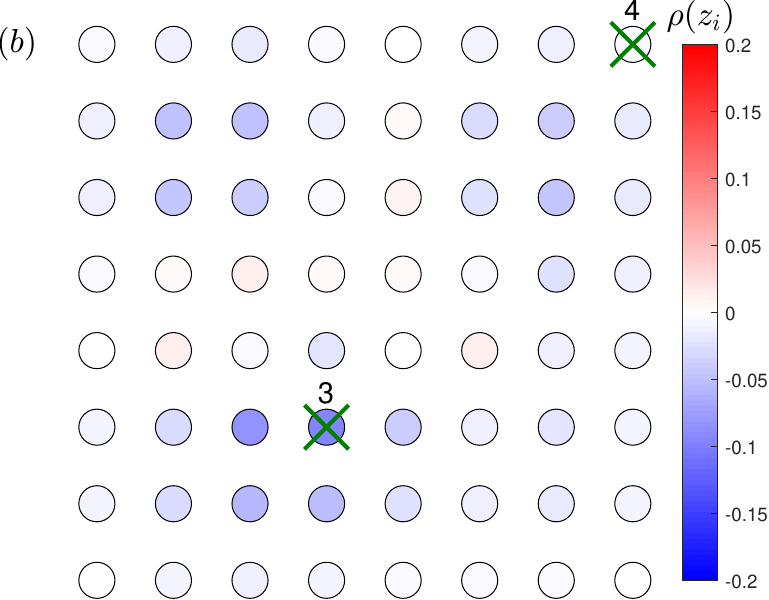}\hfill
\includegraphics[width=0.3\linewidth]{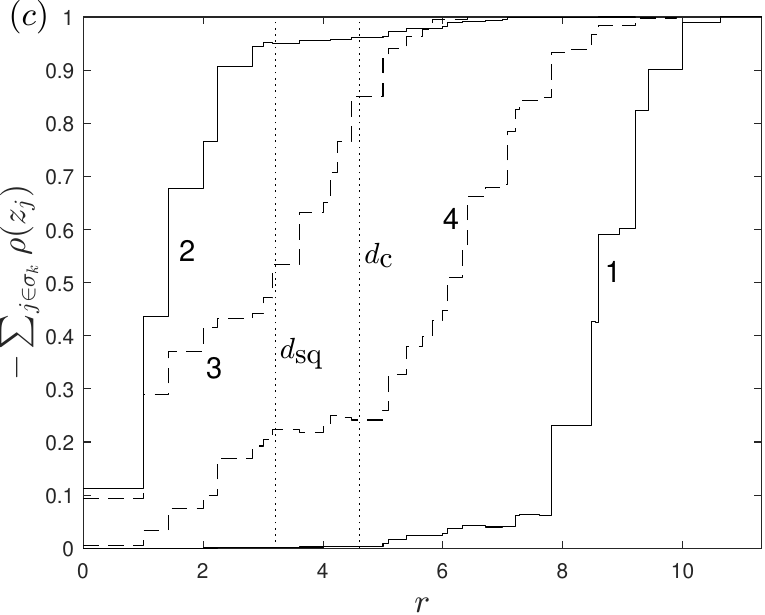}}
\caption{We plot $\rho(z_i)=\langle n_i\rangle_{H+H_V,M-1}-\langle n_i\rangle_{H,M}$ (see Eq.\ \eqref{density}) for (a) the carpet and (b) the square lattice. Here, $M=4$ and potentials with strength $100$ are placed on the two sites marked by green crosses. For the carpet, almost all the charge accumulates in the vicinity of one potential. For the square lattice, the charge is more spread out, but again there is more charge close to one potential than to the other. (c) To quantify this further, we plot the total charge $-\sum_{j\in\sigma_k}\rho(z_j)$ within a circular region with radius $r$ and center at one of the sites on which a potential is applied. Specifically, $\sigma_k$ is the set of all $j$ for which $|z_j-w_k|\leq r$, and $w_k$ is the position of the site labeled $k\in\{1,2,3,4\}$ in either (a) or (b). The vertical dotted lines show half the distance between the two sites on which the potentials are applied for the square lattice ($d_{\textrm{sq}}$) and the carpet ($d_{\textrm{c}}$). If the potentials had trapped two well-separated anyons of charge $1/2$, the curves in (c) would have had plateaus at $0.5$ for $r$ large compared to the size of the anyons and short compared to the distance between the potentials. One should, however, be cautious in interpreting (c) for small lattices, as the trapped charges could have shapes far from circular. Nevertheless, considering (a) and (c) together shows that the potentials do not trap two well-separated anyons on the 64-site carpet with 3 particles, and (b) and (c) suggest that separated anyons are also not trapped on the 64-site square lattice with 3 particles.}\label{fig:density}
\end{figure*}

\textit{Search for topology}---%
One way to demonstrate the presence of topological order is to create anyons and show that they have anyonic braiding statistics. As quasiholes give rise to a local reduction in the density, local potentials tend to trap quasiholes, and braiding can be done by moving the potentials adiabatically \cite{Mueller1}. To compute the statistics of two Abelian anyons, one first computes the phase acquired by the wavefunction when the anyons are exchanged, and then one subtracts the Aharonov-Bohm phase, which is the phase acquired by the wavefunction when one anyon moves along the closed path defined by the exchange, while the other anyon is far away. If the exchange path encloses no area, the Aharonov-Bohm phase is zero, and one can skip the latter step. Quasiholes have a finite spread, and they should be sufficiently far apart that they are well-separated throughout the braiding operation.

We first consider the trapping step. If the considered model is topological, we expect the quasiholes to have charge $1/2$ as mentioned above, and each of them hence reduce the local density by $1/2$ particle. We therefore reduce the number of particles on the lattice by one and add two local potentials
\begin{equation}\label{Hv}
H_V = Vn_l + Vn_m, \quad l \neq m, \quad U \gg V\gg J,
\end{equation}
to the Hamiltonian. We consider the density profile
\begin{equation}\label{density}
\rho(z_i) = \langle n_i \rangle_{H+H_V,M-1} - \langle n_i \rangle_{H,M},
\end{equation}
which we define as the difference between the particle density for the ground state of $H+H_V$ with $M-1$ particles and the particle density for the ground state of $H$ with $M$ particles. If the model is topological, we expect
\begin{equation}\label{charge}
Q_k = - \sum_{z_i\in \sigma_{k}} \rho(z_i), \quad k\in\{1,2\},
\end{equation}
to be $1/2$ if the region $\sigma_k$ is large enough to enclose the $k$th quasihole, but small enough to not enclose the other quasihole. We show one example in Fig.\ \ref{DensityChargeSqSC}(b-c), where we take $\sigma_k$ to be the sites inside the dashed circles. For this example, $Q_1=Q_2=0.466$ for the square lattice and $Q_1=Q_2=0.480$ for the fractal lattice, which are close to the expected value $1/2$. Note that $Q_1=Q_2$ in both cases due to symmetry. To compute the Aharonov-Bohm phase, however, we need to take one anyon far away from the exchange path, and we therefore also consider the case, where one anyon is placed on a corner site. In this case $Q_1$ and $Q_2$ are far from $1/2$ as seen in Fig.\ \ref{fig:density}. This is due to the ground state of the Hamiltonian without potentials and $M$ particles having no significant particle density in the corner site. Meaning that even though a large pinning potential is used, there will be almost no density there to be impacted by the potential. This shows that we should avoid approaching the edges of the lattice too closely, and we shall therefore below consider an exchange path that encloses no area to eliminate the need to compute the Aharonov-Bohm phase.

We next explain the computation of the exchange statistics in further details. The exchange of the pinning potentials is done in the counterclockwise direction. This exchange results in the ground state $|\Psi\rangle$ of the Hamiltonian $H+H_V$ acquiring a Berry phase $\exp(i\pi \theta)$, defined by
\begin{equation}\label{Berry}
\theta =  i \oint_{\mathcal{C}}  \langle \Psi | \nabla_w | \Psi \rangle d w + \text{c.c.},
\end{equation}
where $w$ parametrizes the exchange path $\mathcal{C}$. In general, there are two contributions to consider in $\theta$; the Aharonov-Bohm phase $\theta_{\text{AB}}$, since the trapped charges circulate around the magnetic fluxes, and the statistical phase $\theta_\text{s}$ itself. Therefore we have $\theta = \theta_{\text{AB}} + \theta_\text{s}$. The particular value of $\theta_\text{s}$ ($\neq [0,1]$) characterizes the type of anyons present in a given topological order.

To adiabatically move a trapping potential from the site $l$ to the nearby site $l'$, we follow the procedure in Ref.\ \cite{Mueller1} and consider the Hamiltonian
\begin{equation}\label{H_A_Braid}
H_T = H + (1-\gamma) V n_l + \gamma Vn_{l^\prime} + V n_{m}.
\end{equation}
We vary $\gamma$ from $0$ to $1$, following the ramp
\begin{equation}
\gamma = \frac{\delta r}{r} - \frac{1}{2 \pi} \sin \Bigg(\frac{2 \pi \delta r}{r}\Bigg),
\end{equation}
with $r$ a number of steps sufficiently large to maintain adiabaticity and $\delta r \in [0,1,\ldots,r]$ as the individual step. We move one trapping potential at a time while keeping the other trapping potential fixed at its position to minimize overlap during the driving. The exact diagonalization used to obtain the ground state in each step does not fix the global phase factor of the state. We hence need to fix the global phase factor relative to the state at the beginning of the adiabatic evolution. We do this by choosing the global phase factor of the ground state in a given step such that its overlap with the ground state at the previous step is real.

\begin{figure*}
\includegraphics[width=\linewidth]{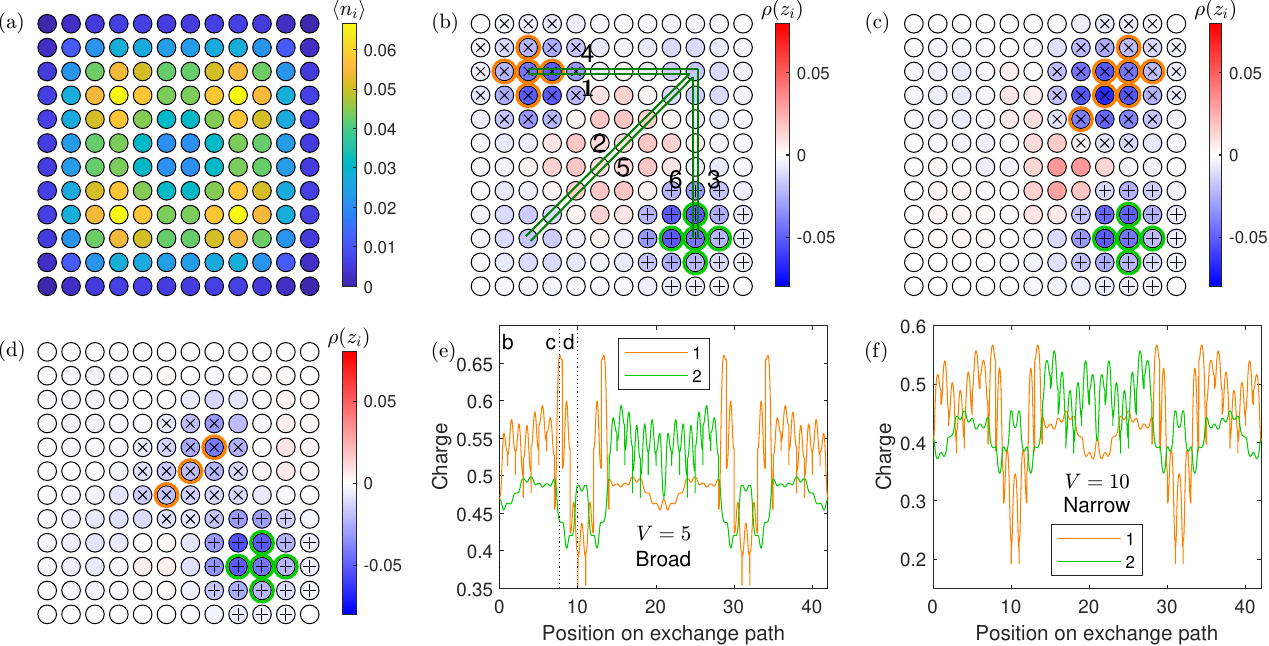}
\caption{(a) Particle density $\langle n_i\rangle$ of the ground state of the Hamiltonian in Eq.\ \eqref{FCI} with $U/J\to\infty$ and $4$ particles on a $12\times12$ square lattice. (b) We exchange potentials along the green paths. To begin with, potentials are placed on the sites $(3,10)$ and $(10,3)$. Potential 1 first moves along path 1 and then path 2, while potential 2 is fixed. Potential 2 then moves along path 3 and 4, while potential 1 is fixed. Finally, potential 1 moves along path 5 and 6, while potential 2 is fixed. The paths are shown slightly displaced to make clear that the exchange happens counterclockwise. Altogether, the path involves moving a potential from one site to the neighboring site $42$ times, and for each such move we use $100$ steps. We hence define the position on the exchange path as a parameter within $[0,42]$ with two decimals. (b-d) show $\rho(z_i)=\langle n_i\rangle_{H+H_V,M-1}-\langle n_i\rangle_{H,M}$ (see Eq.\ \eqref{density}) for different positions on the exchange path for the case of broad potentials and $V=5$. Sites on which potentials are applied are marked with an additional green or orange ring. The local regions defined in the text are marked by crosses (region 1) and pluses (region 2). (e) shows the charge within the local regions 1 (orange) and 2 (green), i.e.\ the sum of minus $\rho(z_i)$ over the sites in the region, for broad potentials and $V=5$. (f) shows the same but for narrow potentials and $V=10$.}\label{fig:Braid}
\end{figure*}

\emph{Square lattice} -- 
We first discuss the model on the square lattice further. Due to the low particle density at the corners and along the edges of the lattice, one needs to put the potentials closer to the center of the lattice to trap sufficiently large charges, but this does not provide sufficient space for separated anyons on the $8\times 8$ lattice as initially considered in Ref.\ \cite{PhysRevA.105.L021302}. We have therefore further developed our exact diagonalization code to be able to consider larger lattices for the same number of particles. The larger lattice also provides further flexibility to make variations to test the robustness of the results.

Specifically, we consider a $12\times 12$ lattice. While this is approximately doubling the number of sites, for the $M=4$ Hamiltonian this increases the Hilbert space dimension from $6.4\times 10^5$ to $1.7\times 10^7$. The low particle density at the corners and edges of the lattice is seen in Fig.\ \ref{fig:Braid}(a). We here choose a path for the exchange of the potentials that encloses no area (see Fig.\ \ref{fig:Braid}(b)), as this eliminates the need to compute the Aharonov-Bohm phase. This was not possible in the 64-site lattice due to limited space in the bulk. We consider two variations: (i) the potentials move from one site $l$ to the next $l'$ following Eq.~\eqref{H_A_Braid} and we refer to this as narrow potentials. (ii) we do the same as (i), except that we put additional potentials of strength $V(1-\gamma)/K_l$ on $K_l$ sites surrounding site $l$ and additional potentials of strength $V\gamma/K_{l'}$ on $K_{l'}$ sites surrounding site $l'$ and we refer to this as broad potentials. We take the $K_l$ (or $K_{l'}$) sites to be the two neighbors along the path. If site $l$ (or $l'$) is one of the four sites at which the (green) path segments end, see Fig.\ \ref{fig:Braid}(b), the $K_l$ (or $K_{l'}$) sites are instead the four nearest neighbors on the lattice. If an additional potential is applied to a site on which a potential is already applied, the potentials add.

To judge whether the potentials produce local density variations of $1/2$, we plot $\rho(z_i)$ (see Eq.\ \eqref{density}) for several points along the path for the case of broad potentials and $V=5$ in Fig.\ \ref{fig:Braid}(b-d). A video of $\rho(z_i)$ for the full exchange is given in the Supplemental Material \cite{sup}. Although density variations are mainly seen close to the potentials, there are also smaller variations quite far from the potentials. To quantify the charge, we show the sum of minus $\rho(z_i)$ over local regions around the potentials in Fig.\ \ref{fig:Braid}(e). The local regions are selected as all sites that are at most $7/(2\sqrt{2})$ lattice spacings away from one of the sites at which a potential is applied for the case of narrow potentials. This ensures that the two local regions (marked by crosses and pluses, respectively, in Figs.\ \ref{fig:Braid}(b-d)) do not overlap. The charges are seen to often deviate by more than 10\% of the ideal value. The most problematic parts are those for which one of the potentials cross the central part of the lattice. This is due to the lower density at the center of the lattice seen in Fig.\ \ref{fig:Braid}(a). Again, however, one should be cautious about interpreting the charges in Fig.\ \ref{fig:Braid}(e), as the shapes of the trapped charges may not fit the chosen local regions. It is more reliable to judge the presence or absence of topology based on the value and robustness of the phase acquired by the wavefunction for different variations of the exchange. We have computed the phase for different choices of the potentials in Tab.\ \ref{thetas}. Some of the results are close to zero, while others are close to one half in units of $\pi$. Comparing the charges obtained for the broad potential and $V=5$ in Fig.\ \ref{fig:Braid}(e) to the charge obtained for the narrow potential and $V=10$ in Fig.\ \ref{fig:Braid}(f), it is also not clear which one is closest to the ideal case for the topological system, although the former might look slightly better as the average of the charges over the path is closer to one half. For the system sizes that we can reach with our exact diagonalization code, we can hence not conclude whether the system is topological or not.

\begin{table}
\begin{tabular}{rp{22mm}cp{22mm}r}
\hline
\hline
$V$ && Potentials && $\theta_s$\\
\hline
 5 && Narrow && -0.08 \\
10 && Narrow && 0.03 \\
12 && Narrow && 0.05 \\
 5 && Broad && 0.32 \\
 8 && Broad && 0.54 \\
10 && Broad && 0.63 \\
\hline
\hline
\end{tabular}
\caption{The ground state wavefunction acquires the phase $\exp(i\pi\theta_s)$, when the potentials are adiabatically exchanged as in Fig.\ \ref{fig:Braid}(b). The table gives $\theta_s$ for different strengths $V$ and different shapes (narrow or broad, see the main text) of the potentials. The number of steps used when discretizing the path is large enough to ensure convergence of $\theta_s$.}\label{thetas}
\end{table}

\begin{figure}
\includegraphics[width=0.8\linewidth]{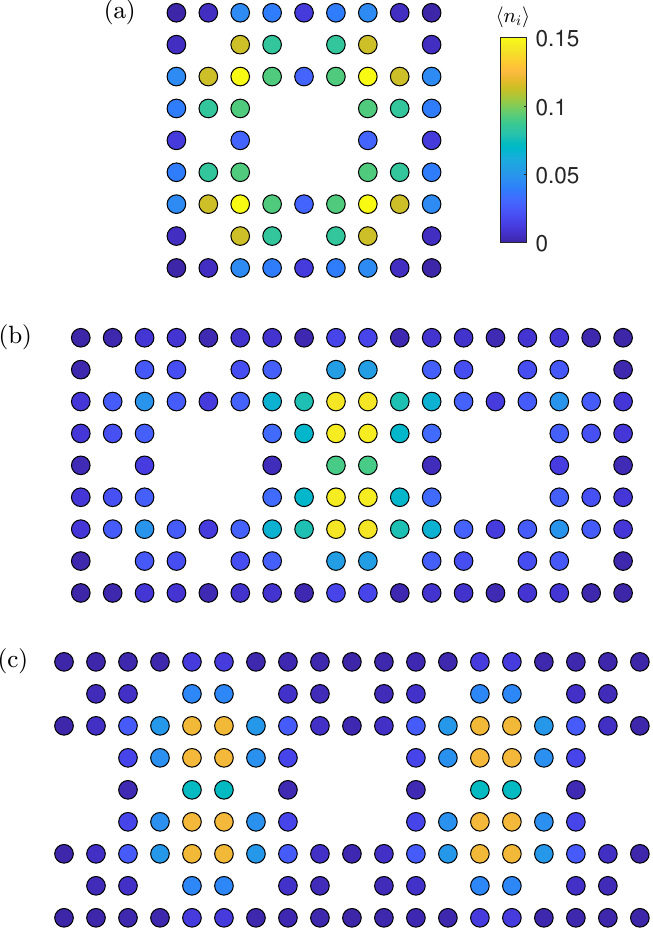}
\caption{Particle density $\langle n_i\rangle$ for the ground state of the Hamiltonian in Eq.\ \eqref{FCI} with $U/J\to\infty$ and $4$ particles for different patches of a high generation Sierpinski carpet with (a) 64 sites, (b) 128 sites, and (c) 130 sites. The highest densities are found in regions that locally look like the two-dimensional square lattice. This conclusion also holds for $M=2$ and $M=3$ particles.}\label{Fig:Carpet}
\end{figure}

\emph{Sierpinski carpet} -- The issues with the low density in the central portion of the square lattice will of course not occur for the Sierpinski carpet, as it has no central bulk. This poses another issue -- how can one generate a braiding path and consistently negate the Aharonov-Bohm phase of that path? We find that this can not be done for any size for which we can realistically calculate the states across the braiding path. This is due to the particle density of the ground state of the $M=4$ local Hamiltonian, which we show for various patches of the Sierpinski carpet of varying size in Fig.~\ref{Fig:Carpet}. The ground state is seen to have high density along quasi-two-dimensional domains. To calculate the statistical phase, paths need to be constructed by moving between multiple quasi-two-dimensional regions (with the local coordination number being four) of sufficient size to support two quasiholes with no overlap, which requires lattices beyond the size capable with the methods implemented here. We note, however, that the very nonuniform density on the patches of the carpet seems inconsistent with the uniform densities observed for the bulk of fractional quantum Hall states on two-dimensional lattices. This gives a hint that the model on the Sierpinski carpet may not be topological.

\textit{Proposal for implementing the Hamiltonian}---%
An experimental demonstration of the Sierpinski carpet fractal lattice in a cold-atom system requires two components: efficient preparation of the lattice system with the desired filling factor and generation of the required site-to-site hopping phases. For the former, we assume that we start by loading a single plane of a 3D cubic lattice in a conventional quantum-gas microscopy system~\cite{Greiner_QGM, Sherson_QGM} capable of imaging atoms with single-site resolution using a high-numerical aperture (NA) microscope objective. Using spin-addressing techniques, a set number of atoms can be loaded into the lattice~\cite{Weitenberg_spin_addressing}.

We now discuss the problem of generating the desired site-to-site hopping magnitudes and phases via light-assisted tunneling~\cite{Aidelsberger_Hofstadter, Greiner_Hofstadter}. In general, tunneling is inhibited in the system if there exists an energy gradient along $x$ and $y$ giving rise to a bias $\Delta$ between each adjacent site. Light-assisted tunneling between adjacent sites can be restored if a pair of running-wave beams is added to the system. Given that the frequency difference between the running waves satisfies the relation $\omega = \omega_1 - \omega_2 = \pm\Delta/\hbar$, atoms are again allowed to tunnel between adjacent sites. The two light fields need only be present where the Wannier functions overlap significantly (that is, between adjacent lattice sites)~\cite{Aidelsberger_review}. Therefore, we can control the magnitude of the effective tunneling parameter through control of the amplitudes of the two running waves we project onto the system. In Refs.~\cite{Aidelsberger_Hofstadter, Greiner_Hofstadter}, the hopping phases were controlled via the relative directions of the two running waves. Here, however, we propose to control the amplitude and phase of the tunneling parameter by locally shaping one of these two running waves using a spatial light modulator (SLM)~\cite{Browaeys_SLM, Greiner_holographic}.

The required tunneling phases are controlled via projection of two counterpropagating light potentials from the top and bottom of the lattice, respectively, with both beams running orthogonal to the lattice axes.
The first laser acts as a light sheet onto the atoms from one direction and does not require high-resolution capabilities. Then, through the high-resolution objective, one can project a second light-based potential with a phase and amplitude pattern mapped onto it via a SLM~\cite{Greiner_holographic}. In this way, one can engineer the local tunneling properties by carefully configuring the system so that light is present only between the adjacent lattice sites where tunneling is desired. If the resolution of the objective is high enough such that the point-spread function of the projection system is comparable to (or smaller than) the distance between lattice sites (see, e.g., the system in Ref.~\cite{Alberti_high_NA}), one can project these light potentials onto the lattice with minimal crosstalk between sites. Even in the presence of small amounts of crosstalk, the SLM-generated light field can readily be modified such that the desired field amplitudes and phases are generated at each lattice site. Finally, given that crosstalk is small, SLMs can also be used to project (again through the high-resolution objective) the local trapping potentials required for anyon generation and exchange.

\textit{Conclusions}---%
We investigated hardcore bosons hopping on square lattices and lattices related to the Sierpinski carpet in the presence of a magnetic field that penetrates the lattice sites. We proposed an experimental implementation of the local model introduced here with ultracold atoms in optical lattices. The experimental realization of the local model studied allows for the consideration of different geometries of the lattice, including different fractal lattices.

Adding trapping potentials and exchanging them adiabatically on a $12\times12$ square lattice along a path that does not enclose an area, the wavefunction acquired a phase that differed by more than one half in units of $\pi$ for different choices of the trapping potentials. From this we concluded that even larger system sizes are needed to judge whether the model is topological or not on a square lattice, but this is beyond the capabilities of the exact diagonalisation methods utilised. For patches of the Sierpinski carpet, we found that the particle density tends to accumulate at regions of the lattice that locally look similar to a two-dimensional square lattice. As fractional quantum Hall states tend to have uniform densities in the bulk, this might suggest that the states on the carpet are not topological.

We have shown that for future explorations of the construction of other local Hamiltonians that could support topological quasiparticles, it is critical that one should aim for ground states with high particle densities across a large connected region of the lattice. This will make the construction of consistent paths to calculate the statistical phase simpler. Future research on which local properties of a Hamiltonian can most impact the extent of quasiparticles and allow for their generation across large regions of two-dimensional lattices would be fruitful.

\begin{acknowledgments}
\textit{Acknowledgements}---The authors thank Blazej Jaworowski for finding an error in a previous version of this manuscript and Wei Wang for discussions on related topics. This work was supported by the Independent Research Fund Denmark under grant number 8049-00074B and by the Carlsberg Foundation through a Semper Ardens grant. Work at the University of Strathclyde was supported by the EPSRC Quantum Technologies Hub for Quantum Computing and simulation (EP/T001062/1). S.M. thanks Weizmann Institute of Science, Israel Deans fellowship through Feinberg Graduate School for financial support.
\end{acknowledgments}

\bibliography{document.bib}

\end{document}